\documentclass[10pt,aps,prl,amsfonts,amsmath,amssymb,raggedbottom,longbibliography,reprint,superscriptaddress,citeautoscript]{revtex4-2}
\usepackage[usenames,dvipsnames]{color}
\usepackage{graphicx,microtype}
\usepackage{textgreek}
\usepackage[bookmarks=false,colorlinks]{hyperref}
\hypersetup{linkcolor=RubineRed,citecolor=Plum,filecolor=Plum,urlcolor=Aquamarine}
\usepackage{xspace}
\newcommand{\kto}{KTaO\textsubscript{3}\xspace}

\begin{document}
	
	\preprint{APS/123-QED}
	
	\title{Ultrafast Suppression of the Ferroelectric Instability in KTaO$_3$}%

	\author{Viktor Krapivin}
	\email{krapivin@stanford.edu}
	\affiliation{Stanford PULSE Institute, SLAC National Accelerator Laboratory, Menlo Park, California 94025, USA}
	\affiliation{Stanford Institute for Materials and Energy Sciences, SLAC National Accelerator Laboratory, Menlo Park, California 94025, USA}
	\affiliation{Department of Applied Physics, Stanford University, Stanford, California 94305, USA}
	\author{Mingqiang Gu}
	\affiliation{Department of Materials Science and Engineering, Northwestern University, Evanston, Illinois 60208, USA}
	\author{D. Hickox-Young}
	\affiliation{Department of Materials Science and Engineering, Northwestern University, Evanston, Illinois 60208, USA}
	\author{S. W. Teitelbaum}
	\affiliation{Stanford PULSE Institute, SLAC National Accelerator Laboratory, Menlo Park, California 94025, USA}
	\affiliation{Stanford Institute for Materials and Energy Sciences, SLAC National Accelerator Laboratory, Menlo Park, California 94025, USA}
	\author{Y. Huang}
	\affiliation{Stanford PULSE Institute, SLAC National Accelerator Laboratory, Menlo Park, California 94025, USA}
	\affiliation{Stanford Institute for Materials and Energy Sciences, SLAC National Accelerator Laboratory, Menlo Park, California 94025, USA}
	\affiliation{Department of Applied Physics, Stanford University, Stanford, California 94305, USA}
	\author{G. de la Pe\~{n}a}
	\affiliation{Stanford PULSE Institute, SLAC National Accelerator Laboratory, Menlo Park, California 94025, USA}
	\affiliation{Stanford Institute for Materials and Energy Sciences, SLAC National Accelerator Laboratory, Menlo Park, California 94025, USA}
	\author{D. Zhu}
	\affiliation{Linac Coherent Light Source, SLAC National Accelerator Laboratory, Menlo Park, California 94025, USA}
	\author{N. Sirica}
	\affiliation{Center for Integrated Nanotechnologies, Los Alamos National Laboratory, Los Alamos, New Mexico 87545, USA}
	\author{M.-C. Lee}
	\affiliation{Center for Integrated Nanotechnologies, Los Alamos National Laboratory, Los Alamos, New Mexico 87545, USA}
	\author{R. P. Prasankumar}
	\affiliation{Center for Integrated Nanotechnologies, Los Alamos National Laboratory, Los Alamos, New Mexico 87545, USA}
	\author{A. Maznev}
	\affiliation{Massachusetts Institute of Technology, Department of Chemistry}
	\author{K. A. Nelson}
	\affiliation{Massachusetts Institute of Technology, Department of Chemistry}
	\author{M. Chollet}
	\affiliation{Linac Coherent Light Source, SLAC National Accelerator Laboratory, Menlo Park, California 94025, USA}
	\author{James M. Rondinelli}
	\affiliation{Department of Materials Science and Engineering, Northwestern University, Evanston, Illinois 60208, USA}
	\affiliation{Stanford Institute for Materials and Energy Sciences, SLAC National Accelerator Laboratory, Menlo Park, California 94025, USA}
	\author{D. A. Reis}
	\affiliation{Stanford PULSE Institute, SLAC National Accelerator Laboratory, Menlo Park, California 94025, USA}
	\affiliation{Stanford Institute for Materials and Energy Sciences, SLAC National Accelerator Laboratory, Menlo Park, California 94025, USA}
	\affiliation{Department of Applied Physics, Stanford University, Stanford, California 94305, USA}
	\author{M. Trigo}
	\affiliation{Stanford PULSE Institute, SLAC National Accelerator Laboratory, Menlo Park, California 94025, USA}

	\date{\today}%

	\begin{abstract} 
		We use an x-ray free-electron laser to study the ultrafast lattice dynamics following above band-gap photoexcitation of the incipient ferroelectric potassium-tantalate, \kto. %
		We use ultrafast near-UV (central wavelength 266\,nm and 50 fs pulse duration) laser light to photoexcite charge carriers across the gap and probe the ultrafast lattice dynamics by recording the x-ray diffuse intensity throughout multiple Brillouin zones using pulses from the Linac Coherent Light Source (LCLS) (central wavelength 1.3\,\AA\, and $< 10$~fs pulse duration). We observe changes in the diffuse intensity that we conclude are associated with a hardening of the soft transverse optical and transverse acoustic phonon branches along  $\Gamma$ to $X$ and $\Gamma$ to $M$. Using ground- and excited-state interatomic force constants from density functional theory (DFT) and assuming the phonon populations can be described by a time-dependent temperature, we fit the quasi-equilibrium thermal diffuse intensity to the experimental time-dependent intensity. We obtain the instantaneous lattice temperature and density of photoexcited charge carriers as a function of time delay. The DFT calculations demonstrate that photoexcitation transfers charge from  oxygen $2p$ derived $\pi$-bonding orbitals to Ta $5d$ derived antibonding orbitals, further suppressing the ferroelectric instability and increasing the stability of the cubic, paraelectric structure.
	\end{abstract}

	\maketitle

	Understanding the interplay between a material structure and its functionality is paramount to devising new technologies. This relationship is evidenced clearly in transition metal oxides (TMO) where multiple strongly interacting degrees of freedom (spin, charge, lattice) give rise to rich phases separated by small energy barriers resulting in giant material responses to external stimuli such as fields or pressure \cite{dagotto2005}. These responses could be exploited for various applications, from energy conversion and light harvesting \cite{Nechache2015,Butler2015} to nanoscale heat management \cite{Luckyanova2012} and information storage\cite{Scott2007,bersukerPseudoJahnTellerOrigin2012,young2015}. 
	Within TMOs, the ABO$_3$ cubic perovskite structure is the building block for many important materials such as ferroelectrics and multiferroics \cite{hillWhyAreThere2000,cheongMultiferroicsMagneticTwist2007,rameshMultiferroicsProgressProspects2007,fiebigEvolutionMultiferroics2016}, with multiple competing phases influenced by particular electronic state configurations  \cite{bhallaPerovskiteStructureReview2000,birolOriginGiantSpinlattice2013,porerUltrafastTransientIncrease2019} as well as anharmonic interactions\cite{young2015}. New strategies for realizing novel phases and functionality can be devised from understanding how microscopic structural and electronic features, e.g., spontaneous polarizations, can be modified by light pulses.
	
	There is increasing interest in manipulating materials using ultrafast pulses to induce novel phases not accessible in equilibrium \cite{basovPropertiesDemandQuantum2017}. In the non-equilibrium state, our understanding of the behavior of coupled electrons and lattice at ultrafast timescales is limited, in part because of the lack of ultrafast atomic-scale probes of the photoexcited material. 
	X-ray free electron lasers (XFEL)s \cite{emmaFirstLasingOperation2010,ishikawaCompactXrayFreeelectron2012,kangHardXrayFreeelectron2017,milneSwissFELSwissXray2017,abeghyanFirstOperationSASE12019} enable probing of coupled electron and lattice degrees of freedom with \AA{}ngstrom wavelengths and at sub-picosecond timescales. 
	Recent experiments using XFELs in EuTiO\textsubscript{3} and doped SrTiO\textsubscript{3} find strong modifications of the interatomic potential upon photoexcitation \cite{porerUltrafastRelaxationDynamics2018,porerUltrafastTransientIncrease2019}.  However, these measurements were limited to the Bragg peak response, and thus provide information about the average crystal unit cell. In contrast, by probing the intensity \emph{between} Bragg peaks with ultrafast x-ray diffuse scattering \cite{trigoFouriertransformInelasticXray2013} we can visualize the evolution of the fluctuations from perfect order. Notably, ultrafast x-ray diffuse scattering is sensitive to nonequilibrium lattice dynamics, including phonons with wavevectors spanning large regions of reciprocal space,  and  thus can be used to obtain the transient interatomic potential and corresponding forces in the photoexcited state \cite{ teitelbaumMeasurementsNonequilibriumInteratomic2021}. This approach provides new insights into structural transitions \cite{Wall2018,jiangOriginIncipientFerroelectricity2016}.

	\begin{figure*}
		\centering
		\includegraphics[width=0.91\linewidth]{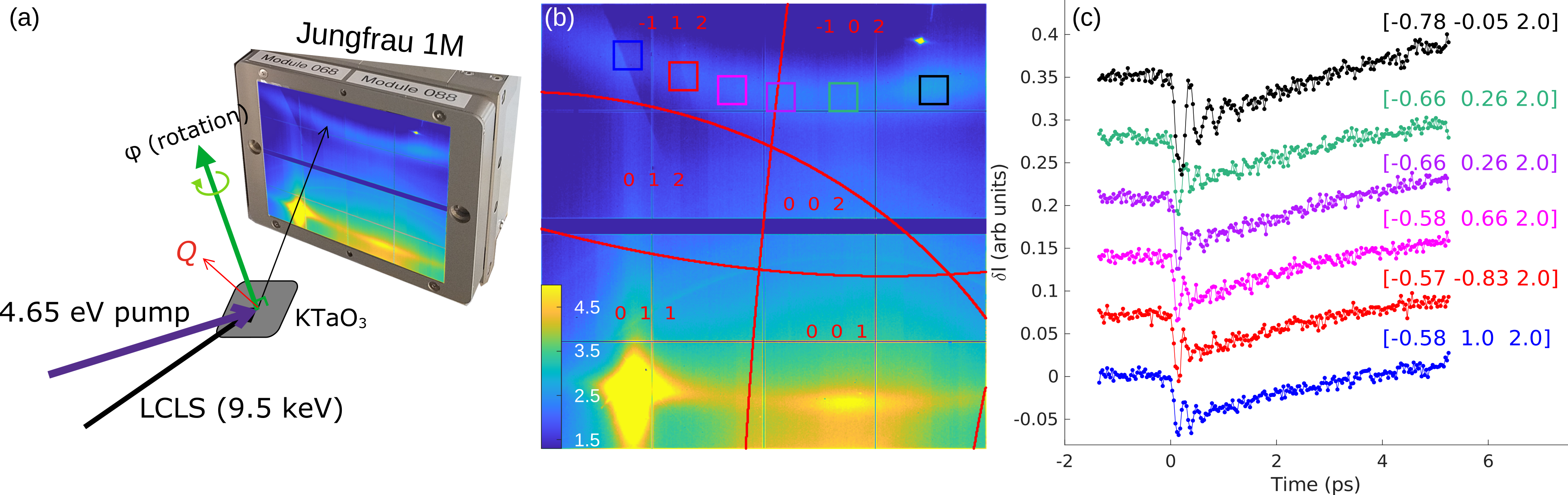}
		\caption{\label{fig:1} (a) Schematic of the grazing incidence scattering geometry. Pump and probe are represented by purple and black arrows, respectively. The scattered x rays are collected with an area detector positioned 110 mm from the sample. The momentum transfer $\mathbf{Q}$ is shown schematically.  The axis of $\phi$ rotation is parallel to the sample normal and $\phi = 0$ corresponds to the $(100)$ direction parallel to the incident x ray beam.
			(b) Room temperature diffuse scattering pattern from \kto for $\phi = 16.98^{\circ}$. The red lines and labels show the boundaries between Brillouin zones and the corresponding reciprocal lattice indices. The scale bar represents linear intensity scale in arbitrary units.
			(c) time dependence of the intensity averaged over each box indicated in (b). The labels indicate the reduced wavevector of the center of each box in reciprocal lattice units.
		}
	\end{figure*}
	
	Here we study KTaO$_3$, which is the structurally simplest member of a broad class of materials based on the ABO$_3$ perovskite structure. This parent structure leads to multitude of instabilities that may involve oxygen octahedra rotations (e.g. SrTiO$_3$), and/or off-centering of the A  or B ions (e.g., BaTiO$_3$ and PbTiO$_3$) \cite{benedekWhyAreThere2013}. The resulting effective anharmonic interaction between these structural distortions may be the key to novel materials with ferroelectric \cite{benedekWhyAreThere2013} or multiferroic orders \cite{hillWhyAreThere2000,bersukerPseudoJahnTellerOrigin2012}.
	Earlier work on the lattice dynamics of \kto used neutron scattering and revealed a joint softening of the lowest (at \textGamma{}) transverse optical (TO) and transverse acoustic (TA) phonon branches at low temperatures along $\Gamma-\mathrm{X}-\mathrm{M}-\Gamma$, suggesting a coupling between these branches. The zone-center TO mode has previously been identified as the ferroelectric soft mode of \kto \cite{comesNeutronScatteringAnalysisLinearDisplacement1972,migoniOriginRamanScattering1976}, which trends to zero as the temperature $T\rightarrow0$\,K. However, the material does not develop a ferroelectric polarization at finite temperature, presumably due to quantum lattice fluctuations \cite{perryPhononDispersionLattice1989}. \kto{} is a quantum paraelectric material similar to SrTiO$_3$ \cite{mullerSrTiIntrinsicQuantum1979}, yet simpler, as \kto{} remains cubic to very low temperatures \cite{tyuninaEvidenceStrainInducedFerroelectric2010}.
	
	We use ultrafast hard x-ray diffuse scattering at the Linac Coherent Light Source (LCLS) x-ray free electron laser  to probe the dynamics of the \kto{} lattice over a wide range of momentum space upon ultrafast above-band-gap photoexcitation with 4.65 eV photons. We treat the lattice with an effective time-dependent temperature after $\sim 1$~ps, and we fit the changes to the quasi-thermal diffuse intensity with the phonons calculated from interatomic forces constants from density functional theory (DFT). The effective force constants between atoms in a $2\times2\times2$ \kto supercell were calculated from DFT using a frozen phonon approach to produce a set of interatomic force constants for each atom pair within this supercell \footnote{%
		Density functional theory (DFT) calculations are performed using Vienna Ab-initio Simulation Package (VASP) \cite{Kresse1996,Joubert1999}. The revised Perdew-Burke-Ernzerhof (PBE) exchange-correlation functional for solids (PBEsol) is used \cite{Perdew2007}. The projector augmented wave (PAW) method is used to treat the core and valence electrons using the following 
		electronic configurations \cite{Blochl1994}: 3s\textsuperscript{2}3p\textsuperscript{6}4s\textsuperscript{1} for K, 5p\textsuperscript{6}6s\textsuperscript{2}5d\textsuperscript{3} for Ta, and 2s\textsuperscript{2}2p\textsuperscript{4} for O. A $12\times12\times12$ \textGamma{}-centered Monkhorst-Pack $k$-point mesh is used for calculating the electron density of states (DOS). 
		Lattice dynamical properties are  computed using the Phonopy code \cite{togoFirstPrinciplesPhonon2015} based on the frozen-phonon method with a $2\times2\times2$ supercell.
	}. We interpret the reduction in diffuse scattering as due to a stiffening of the low frequency transverse acoustic (TA) mode, which is associated with a further stabilization of the cubic phase away from the incipient ferroelectric state.  In combination with DFT, we relate the observed changes in the interatomic forces with photoexcitation of bands with $\pi$-bonding (electrons) and anti-bonding (holes) character between the apical oxygen 2$p(x)$ and Ta 5$d(xz)$, and $p(y)$ and Ta 5$d(yz)$.

	The laser-pump, x ray-probe experiment was conducted at the LCLS X-ray Pump Probe (XPP) experimental station \cite{cholletXrayPumpProbe2015,emmaFirstLasingOperation2010,bostedtLinacCoherentLight2016} using x ray pulses $< 10$ fs in duration with an photon energy of 9.5 keV. The x ray beam was incident at a grazing angle of 0.6 -- 0.7 degree with respect to the sample surface. The pump pulses with central energy of 4.65 eV and 50 fs pulse duration were derived from the third harmonic of a Ti:sapphire regenerative amplifier and were incident to the sample at 2.3 degrees with respect to the surface. The pump and probe spot sizes on the sample surface were $0.2\times2.5$\,mm\textsuperscript{2} and $0.2\times1.6$\,mm\textsuperscript{2}, respectively. The pump was p-polarized with an incident fluence of 1.2 mJ/cm$^2$. The x rays scattered from the sample were measured with an two-dimensional detector (Jungfrau Detector, 1024 by 512 pixels, with a 75 $\mu\mathrm{m}$ pixel size \cite{jungmann-smithJUNGFRAUPrototypeCharacterization2014}) positioned 110 mm behind the sample, as depicted in \autoref{fig:1}a. A 2~mm-thick polycarbonate filter was placed in front of the detector to absorb the Ti, K fluorescence at 4.5 keV. The sample was mounted on a rotation stage with the axis of rotation along the sample normal such that $\phi =0$ rotation corresponds to the sample [100]-axis being parallel to the incident x ray beam at $0^\circ$ incidence angle. The arrival time of the probe relative to the pump was measured for each shot \cite{harmandAchievingFewfemtosecondTimesorting2013}, and the data was then sorted based on the corrected pump-probe delay and binned into 30 fs steps. The  scattering patterns in each bin are then summed and are normalized by the total incident intensity in that bin. 
	
	\autoref{fig:1}b shows the static x-ray diffuse scattering pattern from a (001)-oriented single crystal of \kto at room temperature. In this geometry the momentum transfer of the detected x rays covers multiple Brillouin zones (BZ)s of the cubic structure, labeled by their corresponding indices in (b). Broad vertical and horizontal bands are apparent in the diffuse intensity; these originate from thermal diffuse scattering \cite{warrenXrayDiffraction1990} from the soft TA  phonon branches in \kto primarily along the \textGamma{} to X and \textGamma{} to M directions \cite{axeAnomalousAcousticDispersion1970, perryPhononDispersionLattice1989}. Additionally, static disorder within the sample, Compton, and air scattering can contribute a slowly-varying (with respect to $\vec{Q}$) diffuse background. \autoref{fig:1}c shows the time-dependence of the intensity integrated over the colored boxes indicated in (b), which span between the ($\bar{1}$ 1 2) and ($\bar{1}$ 0 2) BZs. The path spanned by these regions of interest (ROI)s approximately follows X to M in reciprocal space. We observe a sudden intensity \emph{decrease} followed by damped oscillations over a wide range of wavevectors. These oscillations appear at twice the frequency of the lowest TA branch, and originate from thermally squeezed phonons modulating the displacement variance at the respective wavevectors~\cite{trigoFouriertransformInelasticXray2013,jiangOriginIncipientFerroelectricity2016}. 
	While these oscillations could be used to extract the frequency of oscillations \cite{jiangOriginIncipientFerroelectricity2016} and, consequentially, the non-equilibrium interatomic forces \cite{teitelbaumMeasurementsNonequilibriumInteratomic2021}, the fast decay in our case results in poor frequency resolution and reduced sensitivity to the forces. Instead, we take a different approach to the same goal and focus on the non-oscillatory dynamics in \autoref{fig:1}c, which can also be related to the interatomic forces \cite{holtDeterminationPhononDispersions1999}. 
	Indeed, the initial decrease in the overall diffuse intensity is unusual, since generally the pump would  increase rather than decrease the diffuse intensity by raising the effective lattice temperature \cite{trigoFouriertransformInelasticXray2013,Wall2018,porerUltrafastRelaxationDynamics2018}. Instead, a \emph{decrease} of intensity is indicative of phonon hardening~\cite{jiangOriginIncipientFerroelectricity2016}.

	\begin{figure}
		\centering
		\includegraphics[width=0.99\columnwidth]{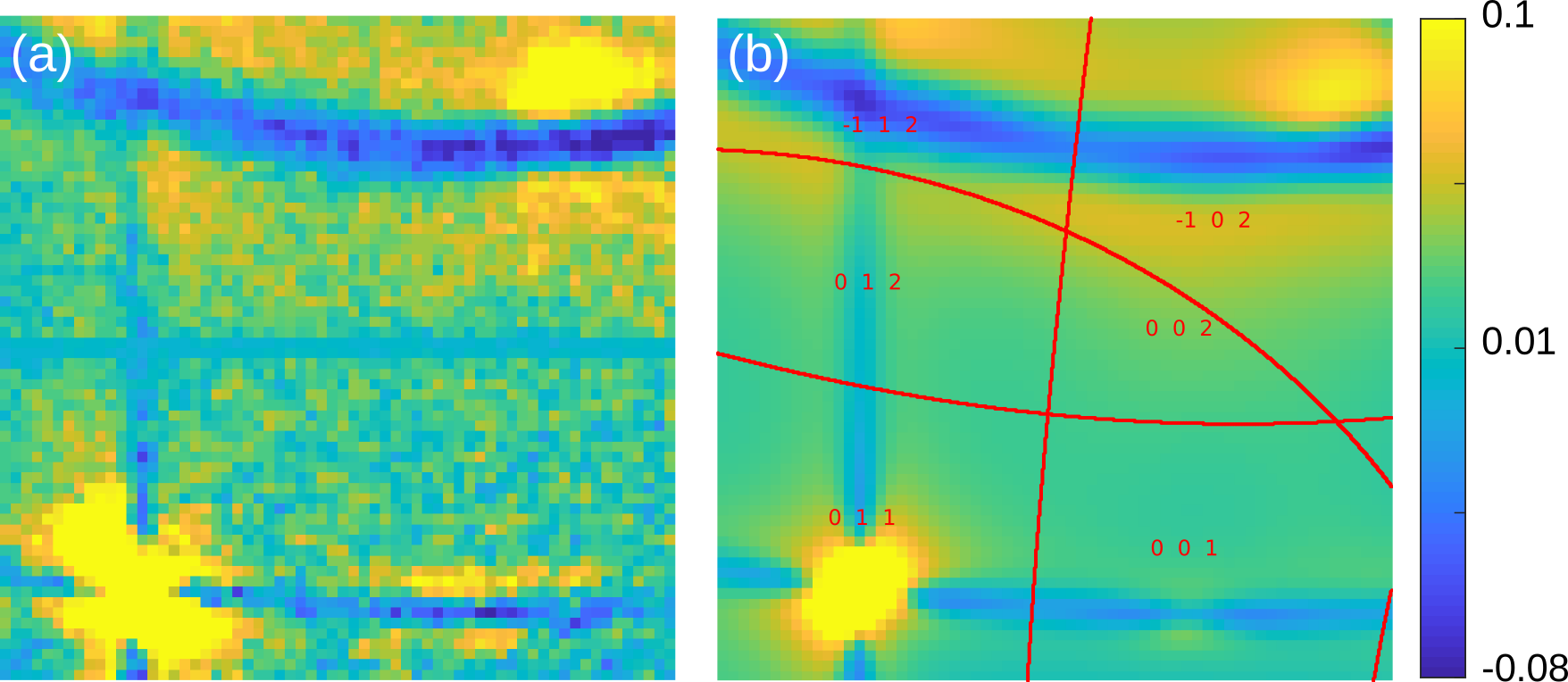}
		\caption{\label{fig:2}
			(a) experimentally measured and (b) fitted $\Delta I(\mathbf{Q},t=0.9\,\mathrm{ps})$. $\Delta I(\mathbf{Q},t)$ in (b) was calculated for $\rho = 0.07$ electrons per unit cell and $T = 430$\,K (see text for details). The scale bar indicates arbitrary units on the same scale as the laser-off image in \autoref{fig:1}.}
	\end{figure}
	
	We begin by extending the equilibrium expression for the thermal diffuse scattering intensity~\cite{warrenXrayDiffraction1990,xuDeterminationPhononDispersion2005} to a time-dependent quasi-equilibrium situation. We approximate the instantaneous (incoherent) phonon populations that give rise to non-oscillatory intensity in \autoref{fig:1} as described by a time-dependent lattice temperature $T$ and instantaneous frequencies. The latter are computed from ab-initio calculations described below. The mode frequencies and displacements are obtained from the eigenvalues and eigenvectors of the (transient) dynamical matrix, which depends parametrically on the density of photoexcited carriers (electrons and holes), $\rho$, that varies slowly with time. Under these assumptions the intensity at a time $t$ and momentum transfer $\mathbf{Q}$,
	\begin{align}
		\label{eq:diffuse}I(\mathbf{Q},t) &=  C \sum_j  \frac{1}{\omega_{j,\mathbf{q}}} \coth \left(\frac{\hbar\omega_{j,\mathbf{q}}}{2 k_B T}\right)     \left| F_{j}(\mathbf{q})\right|^2\\
		F_j(\mathbf{q}) &= \sum_s \frac{f_s}{\sqrt{\mu_s}} \exp(-M_s) (\mathbf{q}\cdot \mathbf{e}_{q_{j,s}})%
	\end{align} 
	$C$ is a constant of proportionality independent of time; $T$ is the time-dependent lattice temperature; $\mathbf{q} = \mathbf{Q}-\mathbf{K}$ is the reduced wavevector, where $\mathbf{K}$ is the closest reciprocal lattice vector to $\mathbf{Q}$; $\omega_{j,\mathbf{q}}$ is the frequency of branch $j$ at wavevector $\mathbf{q}$, $\mathbf{e}_{q,j,s}$ is the eigenvector component of the $s$-th atom in the unit cell for the $j$-th vibrational mode at $\mathbf{q}$. $M_s$ is the Debye-Waller factor, $\mu_s$ is the mass and $f_s$ is the atomic scattering factor of the $s$-th atom in a unit cell, respectively. The quantities  $\omega_{j,\mathbf{q}}$ and $\mathbf{e}_{q,j,s}$ are implicitly functions of $\rho$ as the interatomic forces are a function of $\rho$ \cite{xuDeterminationPhononDispersion2005,warrenXrayDiffraction1990,trigoFouriertransformInelasticXray2013}. Finally, to compare with experiment we consider the change in diffuse intensity
	\begin{align}
		\label{eq:delI}\Delta I(\mathbf{Q},t) &= I(\mathbf{Q},t) - I(\mathbf{Q},t<0)\text{,}
	\end{align}
	where $I(\mathbf{Q},t<0)$ is the x ray intensity recorded with the pump arriving after the x ray probe, which is nearly indistinguishable from the equilibrium diffuse pattern in \autoref{fig:1} (b).

\begin{figure}
		\includegraphics[width=0.99\linewidth]{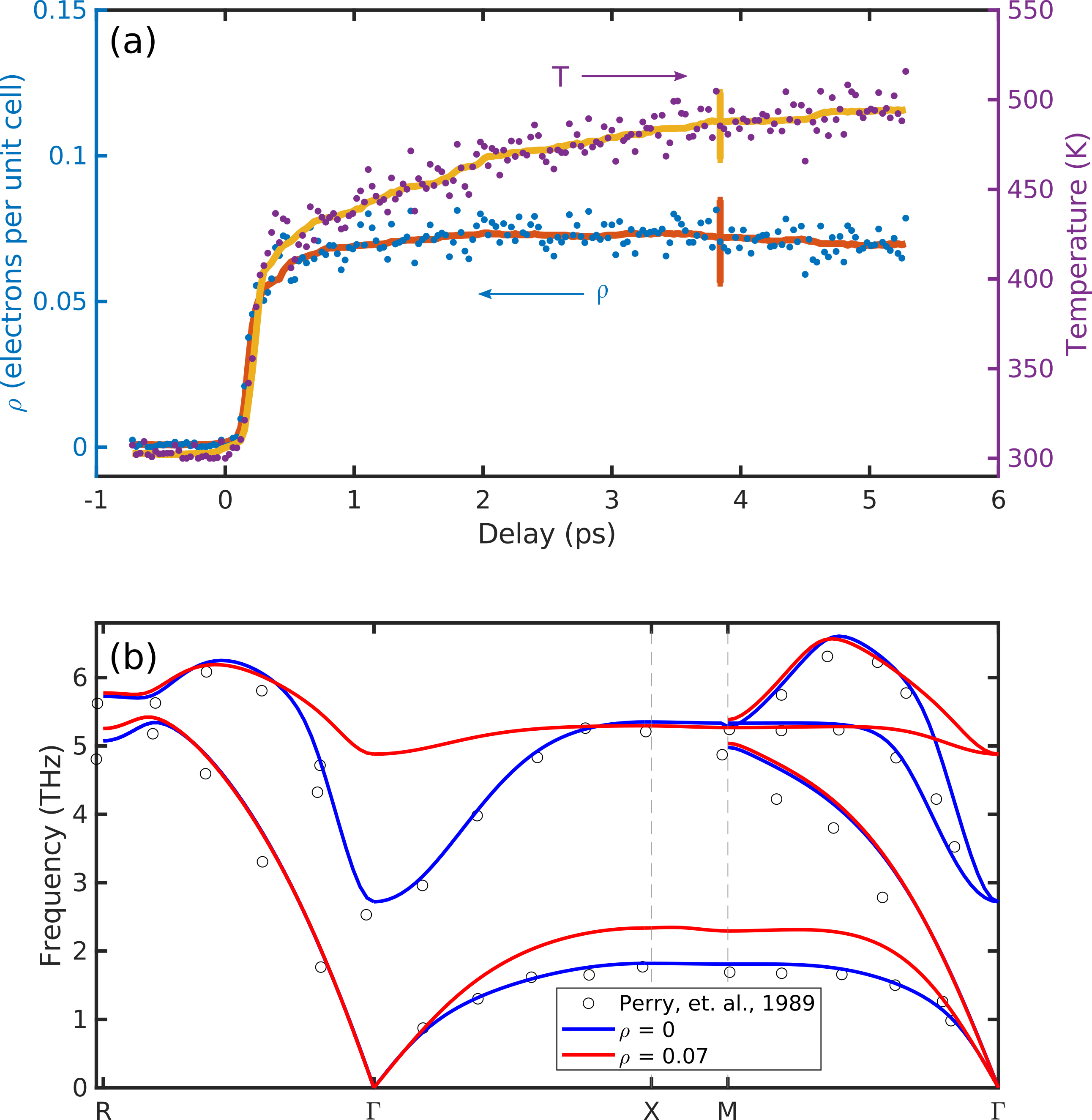}
		\caption{\label{fig:3} (a) fitted carrier density, $\rho$ and lattice temperature, $T$, for each time delay as described in the text. The error bars are representative errors obtained from approximating the covariance matrix from the approximation of the Hessian as $J^T J$, where $J$ is the Jacobian, see \cite{bardNonlinearParameterEstimation1974,doviRemarksUseInverse1991} for details and limitations of this approach. The solid lines show the median value within 30 points.
			(b) The low frequency part of the calculated equilibrium ($\rho = 0$) transverse phonon dispersion of \kto from DFT (blue line). Along the M to $\Gamma$ direction the transverse branch polarized parallel to (1 $\bar{1}$ 0) and polarized parallel to (0 0 1) are plotted. Open circles represent neutron scattering data from \cite{perryPhononDispersionLattice1989}. Red trace: transient dispersion of the lowest TA/TO branches at $\rho = 0.07$ per unit cell, obtained from the fit after 1 ps.}
	\end{figure}

	While a rapid increase in the diffuse intensity may be a signature of an increase in the phonon population (described by the temperature in our approximation), the fast $\sim 100$~fs initial decrease in the intensity in \autoref{fig:1} (c) is too fast to be sudden cooling of the lattice. Instead, we attribute the decrease in the intensity to an \emph{increase} in the phonon frequency [(\autoref{eq:diffuse})] caused by a modification of the interatomic forces by the photoexcited carrier density $\rho$. This is consistent with the initial phase of the oscillations. Thus, we describe the dynamics shown in \autoref{fig:1} (c) using (\autoref{eq:diffuse}-\autoref{eq:delI}) where we allow the temperature to vary with time and the phonon frequencies are obtained from ab-initio density functional theory (DFT), where we assume an instantaneous $\rho(t)$ at each time point. The photoexcited state is approximated in the DFT calculation by constraining the density of electrons (holes) in the conduction (valence) band, $\rho$ \cite{tangneyDensityfunctionalTheoryApproach2002}. Changes in $\rho$ modify the interatomic forces and the corresponding dynamical matrix, from which we obtain the frequencies and eigenvectors in Eq. (\autoref{eq:diffuse}). The dynamical matrix is obtained from force constants computed in a $2\times2\times2$ supercell for $\rho = 0, 0.05$ and $0.10$. The forces were interpolated linearly between these values of $\rho$. We found that the calculated intensity for $\rho = 0$ is in good agreement with the measured equilibrium diffuse intensity \autoref{fig:1}(b) and the known phonon dispersion for the TA and lowest TO branch of \kto \cite{perryPhononDispersionLattice1989}, as shown in \autoref{fig:3}(b) (blue curve). Since the dominant contribution to the diffuse intensity is from low frequency TA modes, here we did not consider LO-TO splitting. We found this does not affect the computed patterns significantly.
	
	As mentioned above, to describe the non-equilibrium diffuse patterns, we assume that $\rho$, $\omega_{{j},\mathbf{q}}$ and $T$ in (\autoref{eq:diffuse}) are time dependent. We then extract $\rho(t)$ and $T(t)$ by fitting (\autoref{eq:diffuse}-\autoref{eq:delI}) to the experimental $\Delta I(\mathbf{Q},t)$ at each time delay with $\rho(t)$, $T(t)$, and the constant $C$ as fitting parameters. To improve the signal to noise and since the relevant features in \autoref{fig:1}b are broad in reciprocal space, we averaged the original images to $64\times64$ pixels. We exclude dead pixels, the region near the (0 1 1) Bragg peak, and the region near the crystal truncation rod visible in the top right of the image in \autoref{fig:1}b as these features do not arise from diffuse scattering, (\autoref{eq:diffuse}), and instead are dominated by strain \cite{reisProbingImpulsiveStrain2001}, coherent oscillations \cite{sokolowski-tintenFemtosecondXrayMeasurement2003} or x ray beam fluctuations. 
	For illustration, the fitted and experimental $\Delta I(\mathbf{Q},t)$ at a single time point $t = 0.9$~ps are shown in \autoref{fig:2}a and b.
	
	The resulting $T(t)$ and $\rho(t)$ are shown in \autoref{fig:3}a. Both quantities quickly rise within $t < 0.5$~ps . Afterwards, $\rho(t)$ saturates while the temperature continues to rise slowly. 
	The agreement shown in \autoref{fig:2} indicates that the phonon population is well parameterized by an effective temperature within the limits of the assumptions in (\autoref{eq:diffuse}) even at relatively short timescales. The interpretation of $T$ as a temperature is more appropriate at later times $t > 1$~ps. The time dependence of $\rho(t)$ results in time-dependent phonon frequencies that affect the slowly-varying diffuse intensity at $t > 1$~ps once the oscillations in \autoref{fig:1}c have decayed.

	In \autoref{fig:3}b, we plot the low frequency region of the phonon dispersion computed for $\rho = 0.07$ ($\rho = 0$) for the lowest two transverse branches in red (blue). 
	At this point it is important to connect the features observed in  \autoref{fig:1}b and \autoref{fig:2}a with the dispersion in  \autoref{fig:3}b. The bright lines forming a square pattern connecting the BZs in \autoref{fig:1}b originate primarily from the soft TA modes along $\Gamma$-X-M shown in blue in \autoref{fig:3}b. Similarly, the $\Delta I(\mathbf{Q},t) > 0$ blue bands in \autoref{fig:2} (a) arise from the hardening of the TA branch along $\Gamma$-X-M shown in \autoref{fig:3} (red curve). Also, note that the prominent hardening of the lowest transverse optical (TO) mode near $\Gamma$ is not observable in this scattering geometry (chosen to avoid the strong elastic scattering of the Bragg peaks at $\Gamma$). 
	We also made a direct estimate of the change in frequency based on the change in intensity at $Q = [-0.77\; 0.00\; 2.01]$ (this location close to the X-point was chosen as the TA phonon branches are nearly degenerate here), which gives $\Delta \omega/\omega$ in the range $ \sim 0.2 - 0.4$  after correcting for the penetration depth mismatch \cite{mamedovKTaO3ReflectionSpectra1984,henkeXRayInteractionsPhotoabsorption1993,jellisonOpticalFunctionsKTaO2006}. This is consistent with the result of the fit obtained in \autoref{fig:3}b, where for a $\rho=0.07$, $\Delta \omega/\omega = 0.2$.

\begin{figure}
		\centering
		\includegraphics[width=0.98\linewidth]{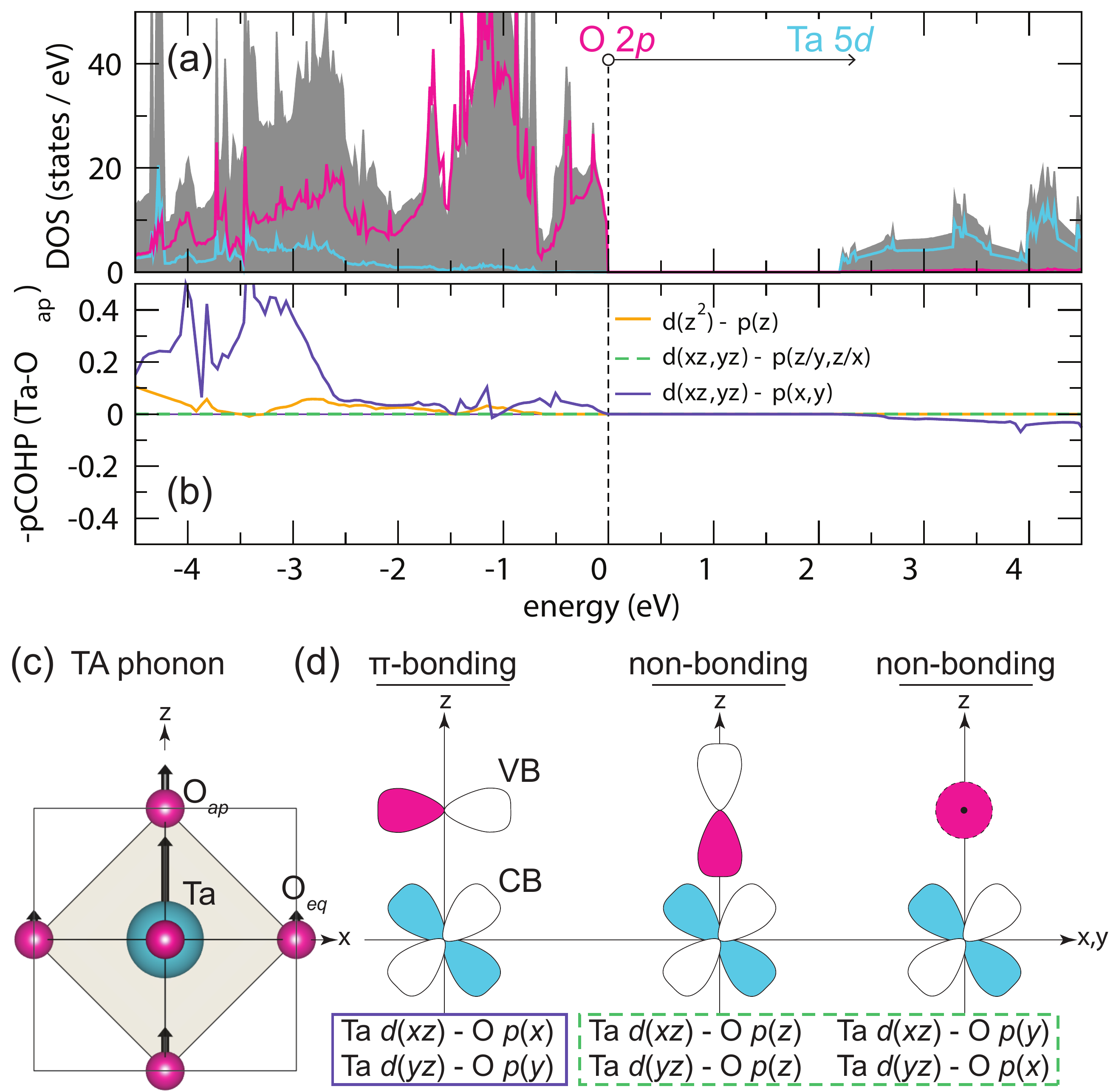}
		\caption{\label{fig:4}
			(a) Calculated electron density-of-states (DOS) and (b) Projected crystal orbital Hamiltonian
			population (pCOHP) for Ta and the apical O atom displacing in the (c) TA mode for KTaO$_3$ at the X point with displacements indicated with black arrows in its electronic ground state.
			The gray filled curve corresponds to the total DOS. The horizontal arrow depicts the photoexcitation of electrons from the bonding O\,$2p$ states at the valence band maximum (VB) to anti-bonding Ta\,$5d$ states forming the conduction band (CB) as illustrated in (b). 
			The photoexcitation  indicated in (a) leads to changes in the (d)  bonding interactions at the valence band maximum  (VBM) to the conduction band minimum (CBM) formed by O\,$2p$ orbitals and Ta\,$5d$ $t_{2g}$ orbitals. 
		}
	\end{figure}

	In the phonon dispersion shown in \autoref{fig:3}b, one of the largest shifts of the TA branch is observed at the X point. This mode corresponds to the anti-polar displacements of the tantalum atoms adjacent to a neighboring unit cell shown in \autoref{fig:4}c. Our DFT calculations also predict that this mode will harden with both $p$ and $n$ doping, irrespective of the charge carriers. 
	We attempted other approximations to photoexcitation within DFT and all approximations resulted in similar qualitative behavior: hardening of the softened phonon dispersion in the photoexcited state 
	\footnote{Three models were applied to approximate photoexcitation within the DFT simulations: (1) Static self-consistent calculations with additional electrons (or holes) compensated by a neutralizing background charge (using Gaussian smearing) followed by frozen-phonon calculations to obtain the IFCs and phonon dispersions. This simulates variable electron and hole concentrations without the corresponding particle that would be photoexcited.
		(2) Simulation of electron-hole pair concentrations: (a) Thermally as described in the main text. Such a simulation is different from the constrained DFT method used in a previous study \cite{paillardPhotoinducedPhaseTransitions2019}; here, we use a large smearing factor (0.6 eV) to close the band gap and enable electron transfer from the VBM to CBM. The TA mode frequency shift is quantitatively consistent to the experimentally observed value when the smearing is $\sim0.8$ eV.
		(b) Athermally by constrained the occupancy of the valence band edge at each $k$-point to compensate for the additional electrons in the conduction band to preserve charge neutrality followed by lattice dynamical calculations\cite{hellmanPotentialenergySurfacesExcited2004}. 
	}.
	
	To further elaborate on the mechanism of the photo-induced phonon mode hardening, consider on changes in the interatomic force matrix elements (IFME) as a function of $\rho$. From DFT, we find the largest change corresponds to a decrease in the nearest neighbor Ta-Ta force constant. This results in a decrease in the corresponding nearest neighbor radial component of the Ta-Ta IFME with respect to $\rho$ by 
	-40 eV/(\AA\,$e$). The second largest IFME change with respect to $\rho$ is an increase between the radial IFME component of the nearest neighbor Ta-O atoms by
	12 eV/(\AA\,$e$). 
	The change in these forces can be understood by examining the electronic structure of \kto from DFT. %
	\autoref{fig:4}a shows the expected insulating behavior in the electronic density of states (DOS).
	The charge transfer gap arises from semicovalent Ta-O interactions that produce 
	a valence band primarily of O $2p$ character separated from a conduction band formed by 
	mainly Ta $5d$ states.
	In the photo-excitation process, electrons are removed from near the valence band edge 
	to near the conduction band edge. 
	\autoref{fig:4}b shows the negative projected crystal orbital Hamiltonian population (-pCOHP) for  
	the apical oxygen ligand with Ta.
	This analysis enables partitioning of the DOS into bonding, nonbonding, and antibonding regions based on orbital overlap matrix elements; bonding, antibonding, and nonbonding interactions exhibit positive, negative, and zero COHP values, respectively.

	We find that the states (derived from the apical oxygen, O$_{ap}$, interacting with Ta) displace in the TA phonon mode at the X-point (\autoref{fig:4}c) and do not significantly contribute bonding character to the valence band edge over the -2.5 to 0\,eV energy range.
	The -2.5 to 0\,eV energy range is dominated by non-bonding interactions (green dashed box in (d)) between Ta-O$_{ap}$ (\autoref{fig:4}d), because the $p(z)/p(y)$ orbitals  approach the $d(xz)$ orbital along its nodes.
	The band edge also consists of $\pi$-bonding interactions between the four symmetry permitted O 2$p(x/y)$  orbitals from equatorial O atoms (O$_{eq}$) with the Ta $d(xy)$ orbitals (not shown). Only two of the O$_{eq}$ atoms weakly participate in the TA mode at the X-point (\autoref{fig:4}c).
	$\sigma$-bonding interactions between Ta $d(z^2)$ and O$_{ap}$ $p(z)$ appear at much lower and higher energy, and do not participate in the photoexcitation process.

	The frequency of the TA branch, therefore, largely respond to occupancy changes in the symmetry permitted $\pi$-bonding (and antibonding) interactions between  O$_{ap}$ $2p(x)$ and  $2p(y)$ orbitals with Ta $5d(xz)$ and $5d(yz)$ orbitals (purple box in \autoref{fig:4}d), which appear from -4.5 to -2.5 eV (and begin at the CBM near $\sim$2.2\,eV).
	Thus, photoexcitation with $4.65$~eV photons depopulate these $\pi$ bonding states and populate antibonding states that control the TA mode stiffness  \cite{bersukerPseudoJahnTellerOrigin2012}.
	The light-controlled occupation of these electronic levels effectively 
	disrupts the O$_{ap}$ bond and reduces the orbital interaction (electron hopping).
	The corresponding IFME then hardens through the vibronic response.

	Our finding give insight into the interaction between symmetry-lowering distortions and the  associated electronic states that are coupled to the fluctuating distortions.
	The dynamic coupling, which is also tuned by the energy separation between the states, governs the stability of TA/TO modes in dielectrics\cite{bersukerPseudoJahnTellerOrigin2012}. 
	For example, ferroelectric compounds that undergo structural transitions stabilized by the aforementioned $p-d$ cross-gap hybridization will exhibit TA/TO modes that harden upon photoexcitation.
	This was suggested in Refs. \cite{paillardPhotoinducedPhaseTransitions2019} for TO modes in ferroelectric perovskite oxides. 
	Our analysis both  accounts for this mode hardening behavior \cite{porerUltrafastTransientIncrease2019,paillardPhotoinducedPhaseTransitions2019} upon photoexcitation and also describes the opposite limit where photoexcitation should affect the vibrational branches weakly, \emph{e.g.}, PbTiO$_3$ and EuTiO$_3$ which have 6s Pb states and Eu 4f in the low energy electronic structure.

	We used ultrafast x-ray diffuse scattering to probe the dynamics of the lattice upon above-gap photoexcitation in \kto. Our analysis of the diffuse intensity across momentum space based on (\autoref{eq:diffuse}-\autoref{eq:delI}) and DFT force constants allowed us to reconstruct the evolution of the transient phonon dispersion and the interatomic forces. We find that photoexcitation induces a hardening of the TA branch suggesting a tendency to move away from the incipient ferroelectric instability. Using ab-initio density functional theory, we find that charge transfer from oxygen p-orbitals to tantalum d-orbitals that form $\pi$-bonding interactions explains the observed changes in the phonon dispersion. The photoexcitation of these $\pi$-bonding states causes the suppression of the Jahn-Teller-like effect and of the ferroelectric instability, which results in the stabilization of the cubic, paraelectric structure of \kto{}. These results suggest that hardening of the TO/TA branches will occur in ferroelectrics with similar $p$-$d$ hybridization, and perhaps less in systems where the $s$ or $f$ orbitals are active.
	
	\begin{acknowledgments}
		
		The experimental work was supported by the U.S. Department of Energy, Office of Science, Office of Basic Energy Sciences through the Division of Materials Sciences and Engineering through FWP No. 2018LANLBES16 (M.-C. L. and R. P. P.), Contract No. DE-AC02-76SF00515 (S. W. T., V. K., Y. H., GdP, M. T. and D. A. R.) and Contract No. DE-SC0019126 (A. M. and K. A. N.). Use of the LCLS was supported by the U.S. Department of Energy, Office of Science, Office of Basic Energy Sciences under Contract No. DE-AC02-76SF00515. Work at Northwestern University was supported by the U.S.\ Department of Energy (DOE), Office of Science, Office of Basic Energy Sciences (BES), under award no.\ DE-SC-0012375. N.S. gratefully acknowledges the support of the US Department of Energy through the Los Alamos National Laboratory LDRD program.
		The computational work used resources of the National Energy Research Scientific Computing Center (NERSC), a U.S.\ DOE Office of Science User Facility located at Lawrence Berkeley National Laboratory, operated under Contract No. DE-AC02-05CH11231, and at the Center for Nanoscale Materials, an Office of Science user facility, supported by the U.S.\ DOE, Office of Science, Office of BES, under Contract No. DE-AC02-06CH11357.
	\end{acknowledgments}

\end{document}